\documentstyle[onecolumn]{mn}

\newif\ifAMStwofonts

\newcommand{\kms}{\hbox{ km\thinspace s$^{-1}$}}    
\newcommand{\arcdeg}{\mbox{$^\circ$}}              

\ifoldfss
  \ifCUPmtlplainloaded \else
    \NewTextAlphabet{textbfit} {cmbxti10} {}
    \NewTextAlphabet{textbfss} {cmssbx10} {}
    \NewMathAlphabet{mathbfit} {cmbxti10} {} 
    \NewMathAlphabet{mathbfss} {cmssbx10} {} 
  \fi
  \ifAMStwofonts
    \ifCUPmtlplainloaded \else
      \NewSymbolFont{upmath} {eurm10}
      \NewSymbolFont{AMSa} {msam10}
      \NewMathSymbol{\upi}     {0}{upmath}{19}
      \NewMathSymbol{\umu}     {0}{upmath}{16}
      \NewMathSymbol{\upartial}{0}{upmath}{40}
      \NewMathSymbol{\leqslant}{3}{AMSa}{36}
      \NewMathSymbol{\geqslant}{3}{AMSa}{3E}

       \let\le=\leqslant
       
    \fi
  \fi
\fi 

\ifnfssone
  \newmathalphabet{\mathit}
  \addtoversion{normal}{\mathit}{cmr}{m}{it}
  \addtoversion{bold}{\mathit}{cmr}{bx}{it}
  \newmathalphabet{\mathbfit} 
  \addtoversion{normal}{\mathbfit}{cmr}{bx}{it}
  \addtoversion{bold}{\mathbfit}{cmr}{bx}{it}
  \newmathalphabet{\mathbfss} 
  \addtoversion{normal}{\mathbfss}{cmss}{bx}{n}
  \addtoversion{bold}{\mathbfss}{cmss}{bx}{n}
  \ifAMStwofonts
    \ifCUPmtlplainloaded \else
      \UseAMStwoboldmath
      \makeatletter
      \new@mathgroup\upmath@group
      \define@mathgroup\mv@normal\upmath@group{eur}{m}{n}
      \define@mathgroup\mv@bold\upmath@group{eur}{b}{n}
      \edef\UPM{\hexnumber\upmath@group}
      \new@mathgroup\amsa@group
      \define@mathgroup\mv@normal\amsa@group{msa}{m}{n}
      \define@mathgroup\mv@bold\amsa@group{msa}{m}{n}
      \edef\AMSa{\hexnumber\amsa@group}
      \makeatother
      \mathchardef\upi="0\UPM19
      \mathchardef\umu="0\UPM16
      \mathchardef\upartial="0\UPM40
      \mathchardef\leqslant="3\AMSa36
      \mathchardef\geqslant="3\AMSa3E

       \let\le=\leqslan

    \fi
  \fi
\fi 

\ifnfsstwo
  \DeclareMathAlphabet{\mathbfit}{OT1}{cmr}{bx}{it}
  \SetMathAlphabet\mathbfit{bold}{OT1}{cmr}{bx}{it}
  \DeclareMathAlphabet{\mathbfss}{OT1}{cmss}{bx}{n}
  \SetMathAlphabet\mathbfss{bold}{OT1}{cmss}{bx}{n}
  \ifAMStwofonts
    \ifCUPmtlplainloaded \else
      \DeclareSymbolFont{UPM}{U}{eur}{m}{n}
      \SetSymbolFont{UPM}{bold}{U}{eur}{b}{n}
      \DeclareSymbolFont{AMSa}{U}{msa}{m}{n}
      \DeclareMathSymbol{\upi}{0}{UPM}{"19}
      \DeclareMathSymbol{\umu}{0}{UPM}{"16}
      \DeclareMathSymbol{\upartial}{0}{UPM}{"40}
      \DeclareMathSymbol{\leqslant}{3}{AMSa}{"36}
      \DeclareMathSymbol{\geqslant}{3}{AMSa}{"3E}

       \let\le=\leqslant

    \fi
  \fi
\fi 

\ifCUPmtlplainloaded \else
  \ifAMStwofonts \else 
    \def\upi{\pi}
    \def\umu{\mu}
    \def\upartial{\partial}
  \fi
\fi

\title[The structure of NGC\,6891]{The Triple-Shell Structure and Collimated 
Outflows of the Planetary Nebula NGC\,6891}

\author[Guerrero et al.]
{Mart\'{\i}n A. Guerrero$^{1,2}$, 
Luis F. Miranda$^3$, 
Arturo Manchado$^{1,4}$, 
Roberto V\'azquez$^{3,5}$ \\
$^1$ Instituto de Astrof\'{\i}sica de Canarias, V\'{\i}a 
     L\'actea s/n, La Laguna 38200, Tenerife, Spain \\
$^2$ Department of Astronomy, University of Illinois at 
     Urbana-Champaign, 1002 West Green Street, Urbana, IL 61801, USA \\
$^3$ Instituto de Astrof\'{\i}sica de Andaluc\'{\i}a, CSIC, 
     Apdo. Postal 3004, 18080 Granada, Spain \\
$^4$ Consejo Superior de Investigaciones Cient\'{\i}ficas, Spain\\
$^5$ Instituto de Astronom\'{\i}a, UNAM, Apdo. Postal 877,
     22800 Ensenada, B.C., M\'{e}xico.
}

\date{Accepted 1999 October 12.
      Received 1999 September 17; 
      in original form 1999 May 12}

\pagerange{\pageref{firstpage}--\pageref{lastpage}}
\pubyear{1999}

\begin{document}
\label{firstpage}

\maketitle


\begin{abstract}

Narrow-band H$\alpha$ and [N~{\sc ii}] images and high-dispersion 
spatially-resolved echelle spectroscopy of the planetary nebula 
NGC\,6891 are presented. 
These observations show a great wealth of structures. 
The bright central nebula is surrounded by an attached shell and a detached 
outer halo. 
Both the inner and intermediate shells can be described as ellipsoids 
with similar major to minor axial ratios, but different spatial 
orientations. 
The kinematical ages of the intermediate shell and halo are $4800$ and 
$28000$ years, respectively. 
The inter-shell time lapse is in good agreement with the evolutionary 
inter-pulse time lapse. 
A highly collimated outflow is observed to protrude from the tips of the 
major axis of the inner nebula and impact on the outer edge of the 
intermediate shell. 
Kinematics and excitation of this outflow provide conclusive evidence that it 
is deflected during the interaction with the outer edge of the intermediate 
shell. 
At the same time, both the kinematics and the morphology of the intermediate 
shell appear to be affected by this interaction.

\end{abstract}



\twocolumn

\section{Introduction}

NGC\,6891 (PN G054.1-12.1) belongs to the very exclusive group 
of triple-shell planetary nebulae (PNe) (Chu, Jacoby, \& Arendt 
1987; Chu et al.\ 1991). 
Its bright, main nebula is surrounded by an attached shell and by an 
additional  detached halo. 
Large, faint haloes in PNe are most likely the result of mass loss 
on the asymptotic giant branch (AGB) during the thermal pulses of 
the helium shell burning phase (TP-AGB). 
The properties of the central star of NGC\,6891 are well known, as it has been 
the object of a detailed analysis using non-LTE models (M\'endez et al.\ 
1988). 
The atmospheric parameters of the central star ($T_{\rm eff}=50,000$~K, 
$\log g=3.9$), its mass (0.75~M$_{\odot}$) and evolutionary stage, as 
well as  the distance to the nebula (3.8 kpc), are known to a high degree 
of reliability. 
Therefore, a kinematical study of NGC\,6891 offers the rare opportunity of 
investigating in an individual PN the multiple-shell phenomenon in relation 
with the final phases in the evolution of its central star (Chu et al.\ 1991; 
Guerrero et al.\ 1996). 
Despite its brightness ($\log~F({\rm H}\beta)=-10.65$, Webster 1983), 
a detailed morphological and kinematical study of the main nebula of 
NGC\,6891 is not yet available. 

Moreover, recent high-dispersion spectroscopy (Guerrero, Villaver \& 
Manchado 1998) suggests the likely presence of a collimated outflow 
in NGC\,6891. 
Collimated outflows are becoming a typical structural component in 
PNe (L\'opez 1997), although  the mechanism for producing collimated 
outflows in PNe is currently an unresolved problem. 
The morphological and kinematical characterization of collimated outflows 
in PNe is of great importance as it is the first step to understanding the 
origin and formation mechanisms of such structures. 

This paper presents high-dispersion spectroscopy and narrow-band 
imaging of NGC\,6891 in the H$\alpha$ and [N~{\sc ii}] $\lambda$6583 
emission lines. 
Special emphasis has been given to the description and analysis of 
the morphology and kinematics of the main nebula and attached shell, 
and to the investigation of the multiple-shell formation. 
The morphological and kinematical properties of the collimated 
outflow and its interplay with the main morphological components 
have also been investigated and reported.

\section{Observations}

Narrow-band CCD images of NGC\,6891 in the H$\alpha$ and [N~{\sc ii}] 
$\lambda$6583 {\AA} emission lines were obtained on 1997 July 25 on the 
2.56-m Nordic Optical Telescope (NOT) at the Roque de los Muchachos 
Observatory (La Palma, Spain). 
The high-resolution adaptive camera (HiRAC) was used in 
combination with a thinned 2k$\times$2k Loral chip, giving 
a plate scale of {0\farcs11} pixel$^{-1}$ and a field of 
view of {3\farcm7}. 
The central wavelength and $FWHM$ of the filters are $\lambda$6563 {\AA} 
and 9 {\AA} in the H$\alpha$ line, and $\lambda$6584 {\AA} and 9 {\AA} 
in the [N~{\sc ii}] line. 
A 15-minute exposure was secured for each filter. 
As deduced from stars in the field, the spatial resolution ($FWHM$) 
is {1\farcs5}. 
The narrow-band images and the [N~{\sc ii}] to H$\alpha$ ratio map are shown 
in Figure~1. 

Long-slit echelle spectra of NGC\,6891 were obtained with the Utrecht 
Echelle Spectrograph (UES) on the 4.2-m WHT at the Roque de 
los Muchachos Observatory on 1995 July 14 and with the IACUB spectrograph 
on the NOT telescope on 1997 July 27. 
The spectral resolution, spatial scale on the detector, and field of view on 
the sky were {6.5\kms}, {0{\farcs}36} pixel$^{-1}$, and {160\arcsec}, for the 
UES spectrum, and {9.5\kms}, {0{\farcs}28} pixel$^{-1}$, and {60\arcsec}, for 
the IACUB spectra. 
Three long-slit spectra were secured at different slit positions; the 
one along the major axis of the main nebula (PA~{315\arcdeg}) was 
observed using UES (1100 s exposure time), whereas the spectra 
at PA~{25\arcdeg} and {260\arcdeg} were collected using IACUB (1800 s 
exposure time). 
The slit positions are shown in Figure~2.

\section{Results}

\subsection{Morphology}

Figure~1 shows the presence of the different shells and morphological 
components in NGC\,6891. 
These are illustrated in Figure~2 in order to provide an easy 
reference. 
The inner shell (the bright main nebula) is not only much brighter in 
the H$\alpha$ image than in the [N~{\sc ii}] one, but its morphology 
is also different. 
The H$\alpha$ image shows a spindle-shaped morphology, $9\arcsec\times6\arcsec$ 
in size, with the major axis at PA~{135\arcdeg} (Fig.~1-{\it top-left}). 
It appears almost rectangular at low intensities (see thick contour in 
Fig.~1-{\it top-left}). 
Two bright emission maxima can be distinguished in the H$\alpha$ image 
along the minor axis (PA~{19\arcdeg}) with a separation of {4\arcsec}. 
It is noticeable that the brightness of the shell edge is point-symmetric, 
the S--SE and N--NW arcs being brighter. 
The morphology closely resembles that of the PN Cn\,3-1 (Miranda et al.\ 1997). 
On the other hand, the [N~{\sc ii}] image (Fig.~1-{\it top-right}) shows 
a remarkably S-shaped morphology. 
The S--SE and N--NW arcs described above in the H$\alpha$ are enhanced in 
the low-ionization [N~{\sc ii}] line. 
The strongest [N~{\sc ii}] emission within these arcs is observed at the 
SE and NW tips of the major axis. 

The intermediate shell is almost circular, with a diameter of {18\arcsec} in 
the H$\alpha$ image (Fig.~1-{\it top-left}). 
The surface brightness profile shows a linear decline as shown in 
Figure~3-{\it left}. 
The [N~{\sc ii}] image departs from the circular shape observed in H$\alpha$ 
but does not follow a well-defined symmetry pattern. 
Two faint knots, A--A$^\prime$, can be observed at PAs~{135\arcdeg} and 
{315\arcdeg} respectively. 
The [N~{\sc ii}]/H$\alpha$ ratio map (Fig.~1-{\it lower-right}) 
reveals two narrow stream-like structures, C--C$^\prime$, which connect 
the major axis of the inner nebula to the A--A$^\prime$ knots. 
The [N~{\sc ii}] emission is also enhanced in two regions, B--B$^\prime$, 
at PAs~{188\arcdeg} and {8\arcdeg}, respectively, where the shell shape 
is swollen (see Fig.~1-{\it top-right}). 
The emission of these regions is more diffuse than in regions A--A$^\prime$.  

The outermost shell of NGC\,6891 is a large ($80\arcsec$ in diameter), faint 
halo detected in H$\alpha$ (Fig.~1-{\it lower-left}), but not in the [N~{\sc 
ii}] line (not shown here). 
The halo is also detected in the [O~{\sc iii}] line (Manchado 
et al.\ 1996). 
High excitation is also observed in other haloes in PNe (Guerrero \& 
Manchado 1999) and may be interpreted as an effect of the {\it hardening} 
of the radiation (Pottasch \& Preite-Martinez 1983), indicating that the 
central regions of multiple-shell PNe (MSPNe) are optically thick to the 
ionizing radiation and that therefore only high-energy UV photons are 
able to reach the halo. 

The halo is almost spherical, but appears slightly distorted towards 
the south where it seems to break and a peculiar spearhead-shaped outer 
structure can be seen. 
The H$\alpha$ surface brightness declines steeply with radius as $r^{-2}$ 
in the inner regions of the halo, but at $r=23\arcsec$ the fall stops and 
a smooth linear decline is observed, as shown in Figure~3-{\it right}. 
The edge of this profile shows a small tip, between radii {34\arcsec} and 
{40\arcsec}, thus indicating a limb-brightened effect that has also been 
reported in other haloes of PNe (Chu et al.\ 1987; Balick et al.\ 1992). 
The profile then declines steeply at low intensities for radii up 
to {40\arcsec}.

\subsection{Kinematics}

Figure~4 shows the long-slit echelle spectra of NGC\,6891. 
The H$\alpha$ line is shown only for the slit at PA~{315\arcdeg}. 
The two emission maxima, located {3\arcsec} from the centre, roughly 
correspond to those observed in the [N~{\sc ii}] line. 
Since the line profile in the H$\alpha$ line is broader than that in the 
[N~{\sc ii}] $\lambda$6583 line, due to its greater thermal broadening, 
the [N~{\sc ii}] line will be used to describe the kinematics of NGC\,6891. 

The long-slit spectra of the inner shell at the different PAs show the 
typical line shape produced by an expanding shell. 
The line spliting is $19.1\pm0.3${\kms} at the central position. 
As measured from the spectra, the shell size is smaller at PAs~{25\arcdeg} 
and {260\arcdeg} (2\farcs0--2\farcs5), close to its minor axis (PA~45\arcdeg), 
than at PA~{315\arcdeg} (the major axis). 
There is no significant ($\le 2$\kms) line tilt at PAs~{25\arcdeg} and 
260\arcdeg, but this increases to $\sim10${\kms} at PA~315\arcdeg. 
Both the morphological and kinematical properties of this shell may be 
described by a prolate shell whose approaching end of the major axis is 
tilted with respect to the line of sight along PA 315\arcdeg. 
A simple ellipsoidal shell model with homologous expansion has been 
used to fit the kinematics of this shell. 
The results are overlaid in Figure~4. 
The best fit is obtained for an inclination of 80\arcdeg. 
The expansion velocity is {17\kms} and {10\kms} for the major and minor axes, 
respectively, and the semi-major axis is {3\farcs8}. 

The point-symmetry morphology of the inner shell is also recognized in 
the echellograms. 
The echellograms at PAs~{25\arcdeg} and {315\arcdeg}, which are placed on 
the S--SE and N--NW brighter arcs, also show a point-symmetric brightness 
enhancement. 
The echellogram at PA~{260\arcdeg}, however, does not show this enhancement 
since this echellogram does not cross the S--SE and N--NW arcs. 

The line shape and brightness of the spectra at PAs~{315\arcdeg} and 
{25\arcdeg} are dominated by two emission maxima at the edge of the 
inner shell that are also observed in the H$\alpha$ line (see above). 
These features extend outwards and can be traced to {9\arcsec} 
from the centre at PAs~{135\arcdeg}--{315\arcdeg}. 
Therefore, the ends of these features at PAs~{135\arcdeg}--{315\arcdeg} 
correspond to the knots A--A$^\prime$. 
Their kinematical structure is better displayed in Figure~5, which shows 
the [N~{\sc ii}] to H$\alpha$ emission line ratio map at PA~315\arcdeg. 
A narrow [N~{\sc ii}]-emission enhanced region (dark grey in Figure~4) 
protrudes from the inner shell edge and crosses the intermediate shell. 
It eventually goes beyond the outer edge of this shell. 
At this position, it increases its velocity by {7\kms} and the emission 
is enhanced in knots A--A$^\prime$. 
The difference in velocity between each pair is only {12\kms} within the 
shell, and {28\kms} at the knots. 
The velocity width of these features, once deconvolved of instrumental 
effects and thermal broadening (assuming a $T_{\rm e}$ of $10^4$~K), is 
very small, between 8 and {10\kms} within the intermediate shell, 
indicating that they are collimated outflows. 
They broaden up to {16--20\kms} at the bright knots A--A$^\prime$. 

The long-slit spectra of the intermediate attached shell also show the 
typical emission line pattern of an expanding shell. 
A line split of {$59\pm1$\kms} is measured at the stellar position. 
The emission line is tilted at PAs~{25\arcdeg} and {315\arcdeg} but 
not at PA~{260\arcdeg}. 
Interestingly, the inner shell emission line at PA~{25\arcdeg} does not 
show the tilt observed for the attached shell. 
This suggests that the symmetry axis of the intermediate shell does not 
coincide with that of the inner shell. 
To confirm this, a model of an expanding ellipsoid was also fitted to the 
data of the intermediate shell. 
The best fit is shown in Figure~4. 
The parameters of the fit are an inclination with the line of sight of 
{50\arcdeg}, an expansion velocity of {$45\pm5$} {\kms} at the poles and 
{28\kms} at the equator, a semi-major axis of {12\farcs0$\pm$1\farcs5} 
and an orientation of the major axis at PA {160\arcdeg$\pm$10\arcdeg}. 
The line shape is very regular at PAs~{25\arcdeg} and {260\arcdeg} which 
closely follow the simple ellipsoidal model used. 
Some small differences are noticeable between the proposed model and the 
kinematical data at PA~{315\arcdeg}, providing evidence that the shell 
shape at this PA departs slightly from an ellipsoid.
In addition, bright condensations are also found at the edge of the 
line at PA~{315\arcdeg}. It is worth noting that they are present as pairs 
showing point-symmetry.

\section{Discussion}

\subsection{The triple-shell structure}

NGC\,6891 is composed of three different shells. 
The bright central part is a prolate shell expanding at $10\kms$ 
at its equator and $17\kms$ at its tips. 
This shell is tilted {80\arcdeg} from the line of sight and oriented at 
PA~{135\arcdeg}. 
The intermediate shell is expanding faster ($28-45\kms$) than 
the inner shell into the outer halo. 
It exhibits the typical linear decline of the emission brightness profile 
of other PN attached shells (Frank, Balick \& Riley 1990; Guerrero et al.\ 
1998). 
The kinematics of this shell suggests that it can also be described by 
an ellipsoidal shell, expanding faster than the inner shell, in agreement 
with the predictions of the current hydrodynamical models (Mellema 1994; 
Steffen et al.\ 1997; Villaver et al.\ 1999). 
The intermediate shell is tilted {50\arcdeg} from the line of sight and 
is orientated at PA~{160\arcdeg}, and therefore the symmetry axes of the 
inner and intermediate shells do not point in the same direction. 
The halo of NGC\,6891 is a faint, almost circular detached shell 
1.5~pc in diameter. 
Its emission brightness falls steeply as $r^{-2}$ until it stops,  
remains almost constant and rises up at the limb-brightened edge. 
Since a constant mass-loss rate would produce an $r^{-3}$ emission 
brightness decline, the profile observed in the inner region of the 
halo reflects either a non-constant mass-loss rate during the late 
AGB evolution, or a redistribution of the material within the shell 
due to dynamical effects, or both. 
The almost constant-brightness region and the tip at the edge of the halo 
strongly indicate that material is piled up at the outer region of the 
halo. 
However, at a height of 800~pc above the Galactic Plane, the hypothesis 
that the interstellar medium (ISM) has significantly retarded the 
expansion of the leading edge of the halo seems implausible. 

Given the distance, angular size and geometry, and expansion velocity of 
each shell in NGC\,6891, the kinematical age may be worked out simply by 
dividing the nebular radius by the expansion velocity. 
Nevertheless, the velocity gradient of {20\kms} from the inner to the 
intermediate shell of NGC\,6891 does not make the direct comparison between 
kinematical and evolutionary ages a straightforward task.
Instead, the expansion velocity and size of the attached shell must be used 
as a reliable estimate of the PN age (Steffen et al.\ 1997). 
At a distance of 3.8~kpc (M\'endez et al.\ 1988), the expansion velocity 
and geometrical parameters of the ellipsoidal expanding shell fitted to 
the intermediate shell in $\S3.2$ results in a kinematical age of $4800$ 
years. 
In contrast, the evolutionary age of the central star, which 
follows from the post-AGB evolutionary tracks of Vassiliadis \& Wood 
(1994) for the central star mass and effective temperature of NGC\,6891 
(M\'endez et al.\ 1998), is too short ($200$ years). 
This would indicate a transition time of $4,600$ years. 

According to the expansion velocity of the halo of {26\kms} reported by 
Guerrero et al.\ (1998), its kinematical age is $28,000$ years. 
The corresponding inter-shell time (the difference between the kinematical 
age of the halo and main nebula) is $23,000$ years. 
This value can be compared with the inter-pulse time, $\tau_{\rm IP}$, based 
on the idea that the haloes and shells of PNe are related to subsequent thermal 
pulses. 
The inter-pulse time can be evaluated using equation 4 from Vassiliadis 
\& Wood (1994) for metallicity Z=0.008 appropriate for high latitude 
Galactic PNe, 
\begin{equation}
\log \tau_{\rm IP} = 7.27 - 3.77\,M_{\rm c},
\end{equation}
where $M_{\rm c}$ is the mass of the PN central star. 
For a core mass of 0.75~$M_{\odot}$ (M\'endez et al.\ 1998), an 
inter-pulse time of $27,700$ years is obtained in good agreement 
with the kinematical inter-shell time. 

\subsection{The low-ionization collimated outflows}

The high collimation degree shown by the gas along C--C$^\prime$ is quite 
noticeable; the measured $FWHM$ of {10\kms} in the [N~{\sc ii}] line 
yields a very small upper limit to the traversal motion of the gas into 
these stream-like structures. 
The maxima velocities along C--C$^\prime$ are measured at the terminal 
knots A--A$^\prime$, which show opposite systemic velocities of 
$\pm~14\kms$. 
Since these velocities are projected on to the sky, the real expansion 
velocity must be faster. 
A lower limit is settled by the expansion velocity of the intermediate 
shell (45\kms), as knots A--A$^\prime$ are overtaking this shell.  
If we assume the inclination angle with the line of sight to be the same 
as the inclination of the inner shell (80\arcdeg), a deprojected 
velocity of {80\kms} is calculated for knots A--A$^\prime$. 
Therefore, C--C$^\prime$ can be interpreted as a fast collimated outflows 
which originate at the inner shell major axis tips and traverse the 
intermediate shell. 

Knots A--A$^\prime$ show strong evidence of the interaction of the 
collimated outflow with the outer edge of the intermediate shell. 
The velocity increases, indicating that the outflow is accelerated or most 
probably deflected at the surface discontinuity between the intermediate 
shell and the halo. 
The $FWHM$ also rises to {20\kms} what indicates that traversal motions and 
turbulence increase in the outflow when it reaches the outer edge of the 
intermediate shell. 
All this indicates that a density enhancement is found at the outer edge 
of this shell, or that the shell itself plays a significant role in the 
collimation of the outflow.
In addition, the line shape at PA~{315\arcdeg} does not exactly fit an 
expanding homologous ellipsoid; instead, emission is enhanced at different 
points in the position--velocity echellogram and the kinematics slightly 
depart from that expected for such model. 
This is not likely to be fortuitous, but strengthens the idea that the 
interaction between the collimated outflow from inwards and the edge of 
this shell is taking place in this nebula. 

Very similar phenomenology has also been described for NGC\,6572 (Miranda 
et al.\ 1999). 
In this case, kinematics and morphology provide conclusive evidence of a 
collimated outflow--shell interaction. 
As a result of this interaction, the shell has been broken up and presents 
peculiar kinematics. 
Collimated outflow--shell interaction can also be described in the case of 
IC\,4593 (Corradi et al.\ 1997; O'Connor et al.\ 1999). 
A detailed inspection of the data by Corradi et al.\ shows that the behaviour 
of the collimated outflows at PAs~{139\arcdeg} and {242\arcdeg} (knots A and 
C respectively as defined by Corradi et al.) is very similar to that 
described for NGC\,6891. 
The outflows at PA~{139\arcdeg} and {242\arcdeg} in IC\,4593 are also sped 
up and the $FWHM$ also increases as it reaches the shell edge at the position 
of IC\,4593 knots A and C (see Fig.~4 of Corradi et al.\ 1997). 
Interaction between collimated outflows and the nebula has been invoked by 
Sahai \& Trauger (1998) to interpret multiple point-symmetric features 
observed at or near the border of the shell in several young PNe. 
All these data strongly suggest that interaction between collimated outflows 
and shells in PNe indeed occurs and may play an important role in the shaping 
of the shell.

The knots A--A$^\prime$, the collimated structures C--C$^\prime$ and 
the regions B--B$^\prime$ present a [N~{\sc ii}] to H$\alpha$ ratio 
enhancement indicative of the presence of low-ionization gas. 
Since they appear as pairs located on opposite sides of the central star, 
they may be considered as FLIERs (``fast low-ionization emission regions''), 
as they share many of the properties described in such structures (Balick 
et al.\ 1998, and references therein). 
Indeed, the morphology and kinematics of NGC\,6891 share many similarities 
with other PNe that present FLIERs. 
In particular, there is a close resemblance to NGC\,3242, NGC\,6826 and 
NGC\,7009, although it must be noted that the collimated outflow described 
in NGC\,6891 is narrower as it is in NGC\,7009. 
NGC\,6751 (Chu et al.\ 1991) and IC\,4593 (Corradi et al.\ 1997; O'Connor 
et al.\ 1999) are also very similar objects. 
All of them show the same structure: an elliptical or prolate inner shell, 
a round attached intermediate shell and a detached halo (Chu et al.\ 1987; 
B\"assgen \& Grewing 1989), and a fast outflow of low-ionization material 
is observed at or near the tip of the major axis of the inner shell 
(Balick et al.\ 1998). 
%

%
%
%
%
%
%
%
%

There is growing evidence for a close relation between FLIERs, 
collimated outflows and ansae, and the inner shells (NGC\,3242, 
NGC\,6751, NGC\,6826 and NGC\,7009) or inner structures (IC\,4593) of 
multiple-shell PNe. 
It is interesting to note that the lowest-ionization regions are 
found in all cases at the outer edge of the intermediate shell. 
These elements seem be of physical importance in the collimation of an 
outflow within the attached outer shell. 
But a new element is added in NGC\,6891: the point-symmetric distribution 
of the brightness of the inner shell, both in the morphology and kinematics, 
may be related to the physical conditions required to the collimating 
mechanism. 

The formation mechanisms of such structures in PNe are still unknown. 
Only under unrealistic conditions for the mass-loss rate (Borkowski, 
Blondin \& Harrington 1997) or restrictive conditions for the fast wind 
velocity (Frank, Balick \& Livio 1996) the proposed models can reproduce 
collimated outflows along the major nebular axis (but see 
Dwarkadas \& Balick 1998). 
The situation turns to be even worse because more exotic phenomena, 
invoking the precession or rotation of the collimating agent,   
are being observed in an increasing number of PNe. 
The precessing bipolar collimated outflows reported for NGC\,6543 (Miranda 
\& Solf 1992) and NGC\,6884 (Miranda, Guerrero \& Torrelles 1999), the 
kinematical properties of point-symmetric PNe (Guerrero, V\'azquez \& 
L\'opez 1999), or the bipolar rotating episodic jets (BRETs) (L\'opez, 
Meaburn \& Palmer 1993; Bryce et al.\ 1997) increase the difficulties of 
the interpretation. 
In such cases, the explanation invokes the presence of a close binary 
companion for the central star of the PN going through a common envelope 
phase, or magnetic fields, or both (see Livio \& Pringle 1997, and 
references therein; Garc\'{\i}a-Segura 1997).  

The great wealth of structures within NGC\,6891 will obviously benefit from 
high-resolution narrow-band images in low-ionization species. 
In particular, these would help to reveal the detailed geometry between 
the inner shell and the collimated outflows C--C$^\prime$, as well as 
to confirm whether the point-symmetry of this shell is connected to the 
collimation mechanism. In addition, intermediate-dispersion spectroscopy of 
the low-ionization emission regions will be very valuable 
for working out the physical conditions and chemical 
abundances in these regions and compare them to other FLIERs.

\section{Summary and Conclusions}

We summarize our results and conclusions below. 

\begin{enumerate}
\item{ The nebular structure of NGC\,6891 reflects the different mass-loss 
       episodes experienced by its progenitor star on the tip of the AGB 
       phase and on the post-AGB stage. 
       The halo was formed some $28,000$ years ago during the mass-loss 
       episode associated with a period between thermal pulses on the late 
       AGB phase. 
       Some $4,800$ years ago, the progenitor star ejected most of its 
       remaining envelope and the intermediate and inner shells arose.
       }
 

\item{ The previous interpretation is in good agreement with the current AGB 
       inter-shell time lapse predictions, although a significant transition 
       time ($4,600$ years) must be allowed. 
       }

\item{ The inner and intermediate shells may be described by a simple 
       model of an expanding ellipsoid. 
       Expansion velocities of 17 and {10\kms} for the inner 
       shell and 45 and {28\kms} for the intermediate shell 
       are worked out. 
       Although the major to minor axis ratio is almost the same for 
       both shells, the spatial orientation of the shells is different. 
       }

\item{ A fast collimated outflow ($v_{\rm exp}\,>\,45\kms$) runs from 
       the tips of the major axis of the inner shell throughout the 
       intermediate shell until it reaches the outer edge of this region. 
       This is interpreted as a FLIER. 
       }

\item{ The kinematics and excitation of knots A--A$^\prime$ provide 
       conclusive evidence of the interaction between the collimated 
       outflow and the outer edge of the intermediate shell. 
       The outflow is deflected during the interaction, while the 
       outer edge of the intermediate shell departs from an elliptical shape. 
       } 
\end{enumerate}

\section*{Acknowledgments}

The William Herschel Telescope and Nordic Optical Telescope are operated 
on the island of La Palma in the Spanish Observatorio del Roque de los 
Muchachos of the Instituto de Astrof\'{\i}sica de Canarias by the Royal 
Greenwich Observatory and Lund Observatory, respectively. 
MAG and AM are partially supported by Spanish grants PB94--1108 and 
PB97--1435--C02--01. 
MAG is also supported partially by the Direcci\'on General de Ense\~nanza 
Superior e Investigaci\'on Cient\'{\i}fica of Spanish Ministerio de 
Educaci\'on y Cultura. 
LFM and RV are partially supported by Spanish grant PB95--0066 
and Junta de Andaluc\'{\i}a. 
RV also acknowledges a graduate scholarship from AECI (Spain) and complementary 
support from DGAPA--UNAM (Mexico). 

\label{lastpage}

\vfill\eject

\vfill\eject

\onecolumn

\clearpage

\begin{figure}
\vspace*{18.0cm}
\includegraphics{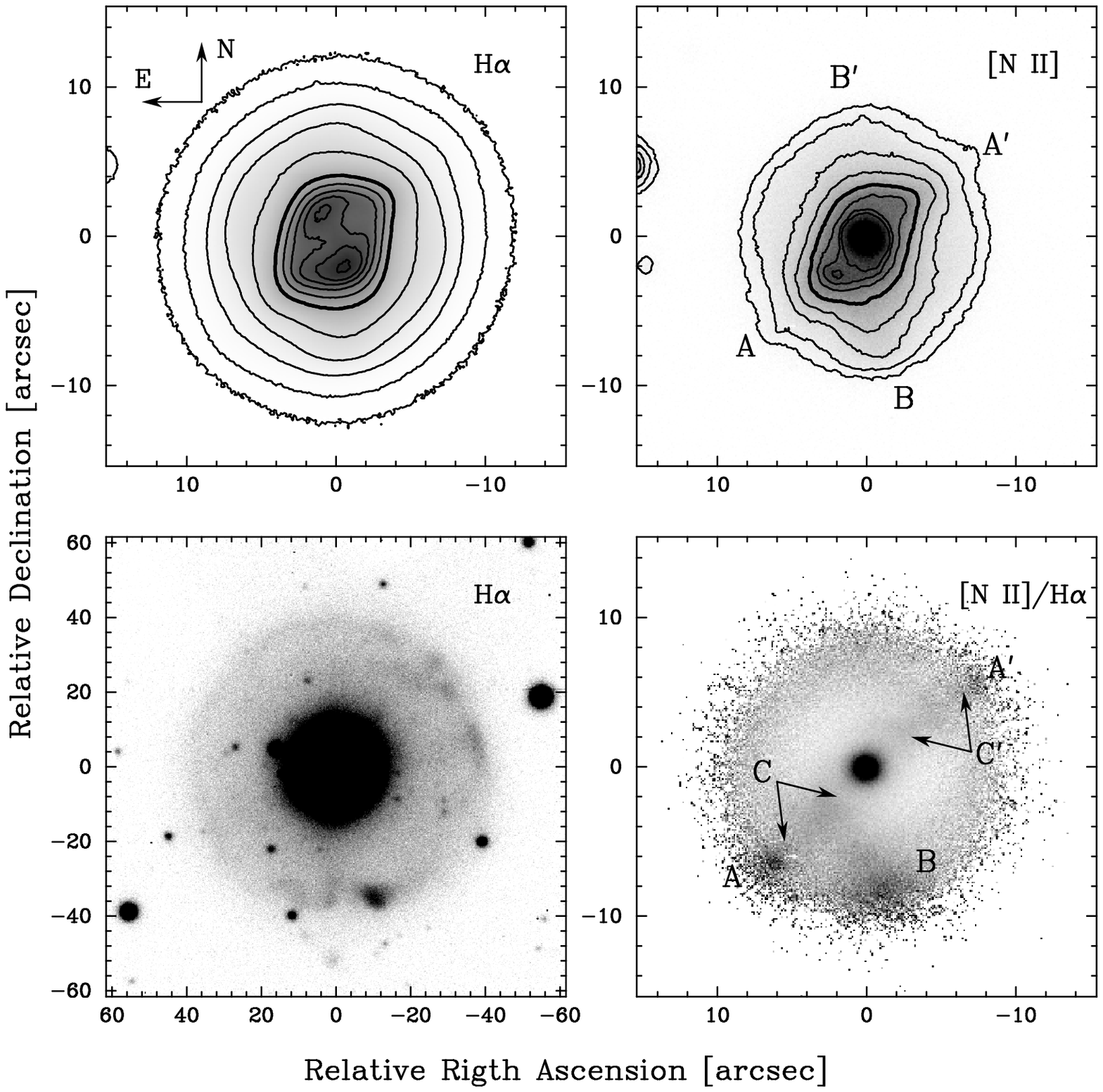}
\end{figure}
     
{\bf Figure~1.--} 
Negative grey-scale representation of CCD images of NGC\,6891 in 
H$\alpha$ ({\it left column}) and [N~{\sc ii}] ({\it top-right}), 
and of the [N~{\sc ii}] to H$\alpha$ ratio map ({\it lower-right}). 
The grey-scale levels have been selected to emphasize the inner 
region an the intermediate shell ({\it top row}), and  
the halo in the H$\alpha$ line ({\it lower-left}). 
The morphological features A--A$^\prime$, B--B$^\prime$ and
C-C$^\prime$ are labelled in the figure. 
Logarithmic contours have been overlaid over the {\it top} figures. 
The thick contour marks the outer limit of the inner shell. 

\clearpage

\begin{figure}
\vspace*{9.0cm}
\includegraphics{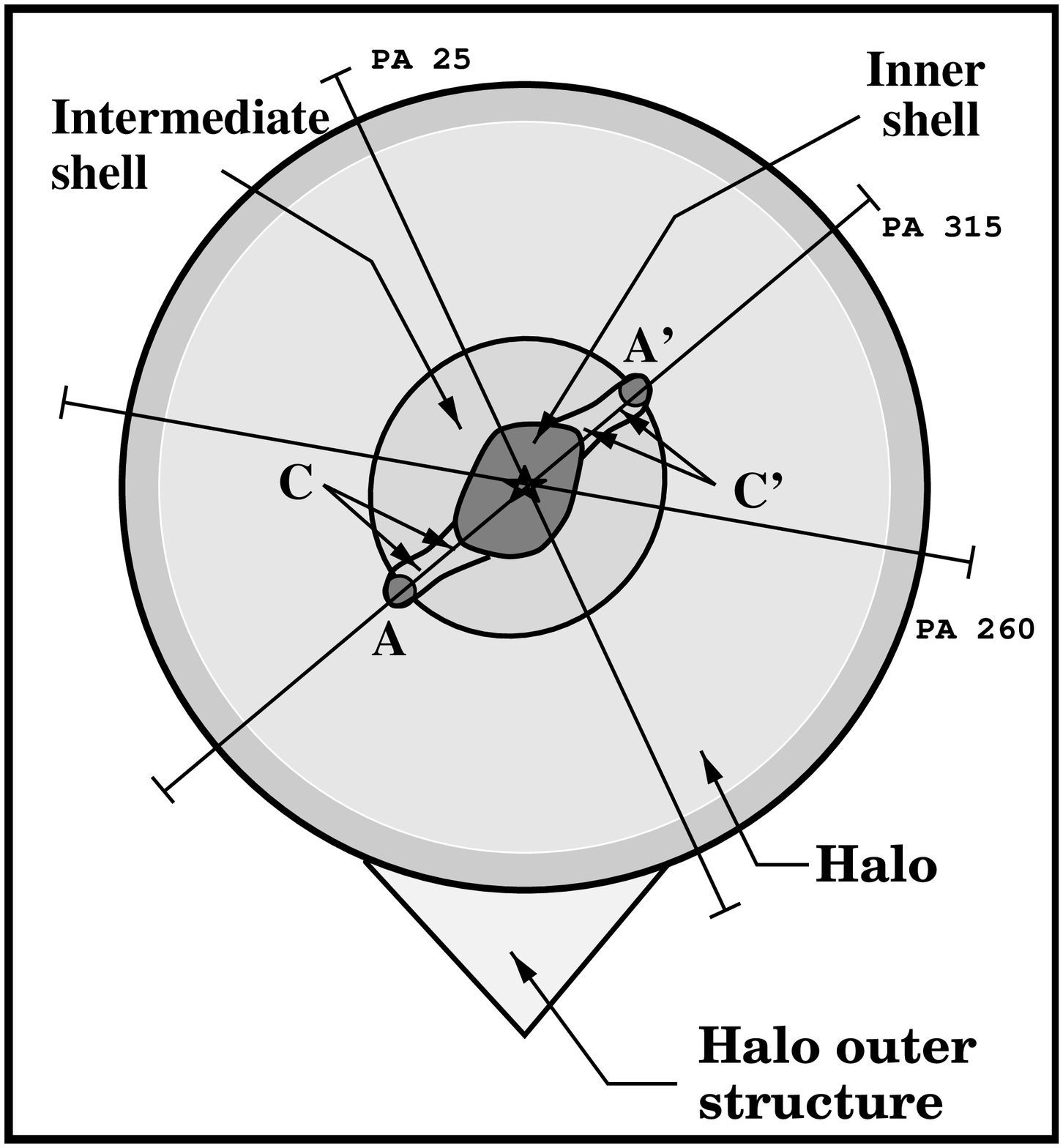}
\end{figure}

{\bf Figure~2.--}
Sketch illustrating the main morphological features observed in 
NGC\,6891 and the position of the slits.

\begin{figure}
\includegraphics{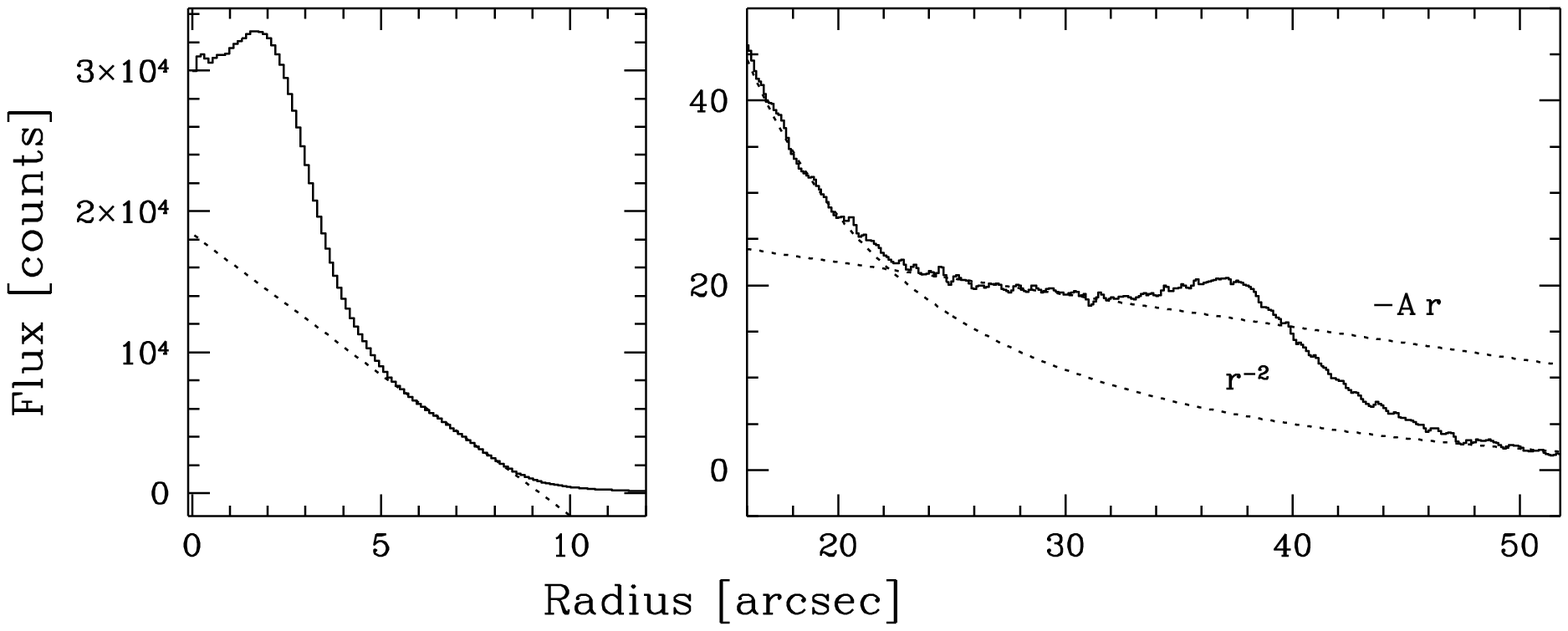}
\end{figure}

\vspace*{8.0cm}
{\bf Figure 3:}
H$\alpha$ surface brightness profile of the inner and intermediate 
shells ({\it left}) and halo ({\it right}) of NGC\,6891. 
The inner and intermediate shells profile was extracted along the 
inner shell minor axis at PA~{45\arcdeg}. 
The intermediate shell is well matched by a linear fit 
(dotted line). 
The halo profile was azimuthally averaged to increase the $S/N$ 
ratio, excluding the southernmost part of the halo (from P.A. 
165\arcdeg to 205\arcdeg) where it is not round. 
The inner region of the halo declines as $r^{-2}$, but the 
region between {23\arcsec} and {34\arcsec} is better 
described by a linear fit ($-A\,r$). 
Both fits are shown (dotted line) in the figure. 

\clearpage

\begin{figure}
\vspace*{9.5cm}
\includegraphics{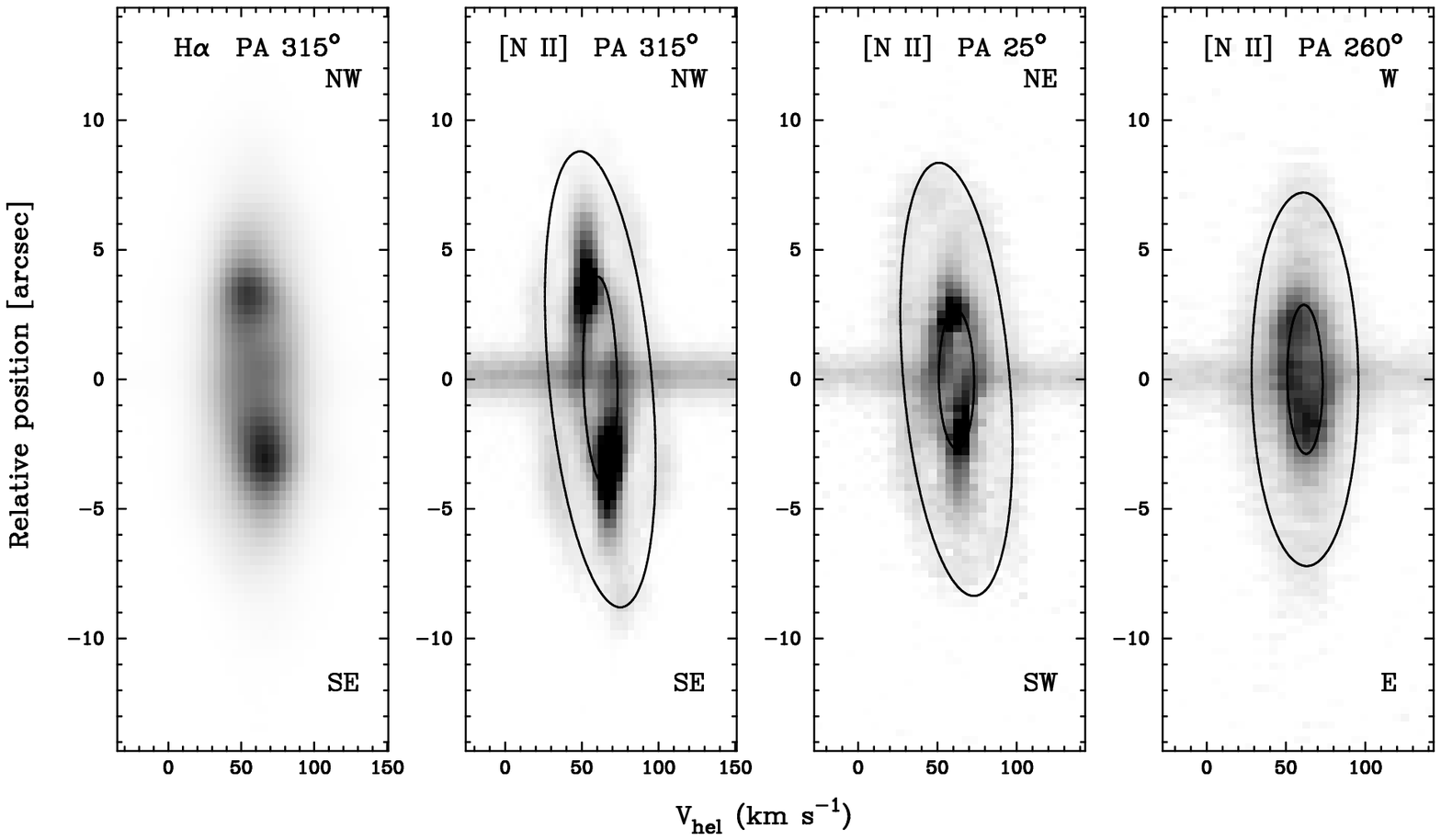}
\end{figure}

{\bf Figure 4.--}
Negative grey-scale representation and contours of the H$\alpha$ 
and [N~{\sc ii}] emission-lines echellograms of NGC\,6891 at 
PA~{315\arcdeg}, {25\arcdeg} and {260\arcdeg}. 
The spatial orientation and scale and the velocity scale are 
given.
The ellipse velocity predictions from the models described in 
the text for the inner and intermediate shells are shown. 

\begin{figure}
\includegraphics{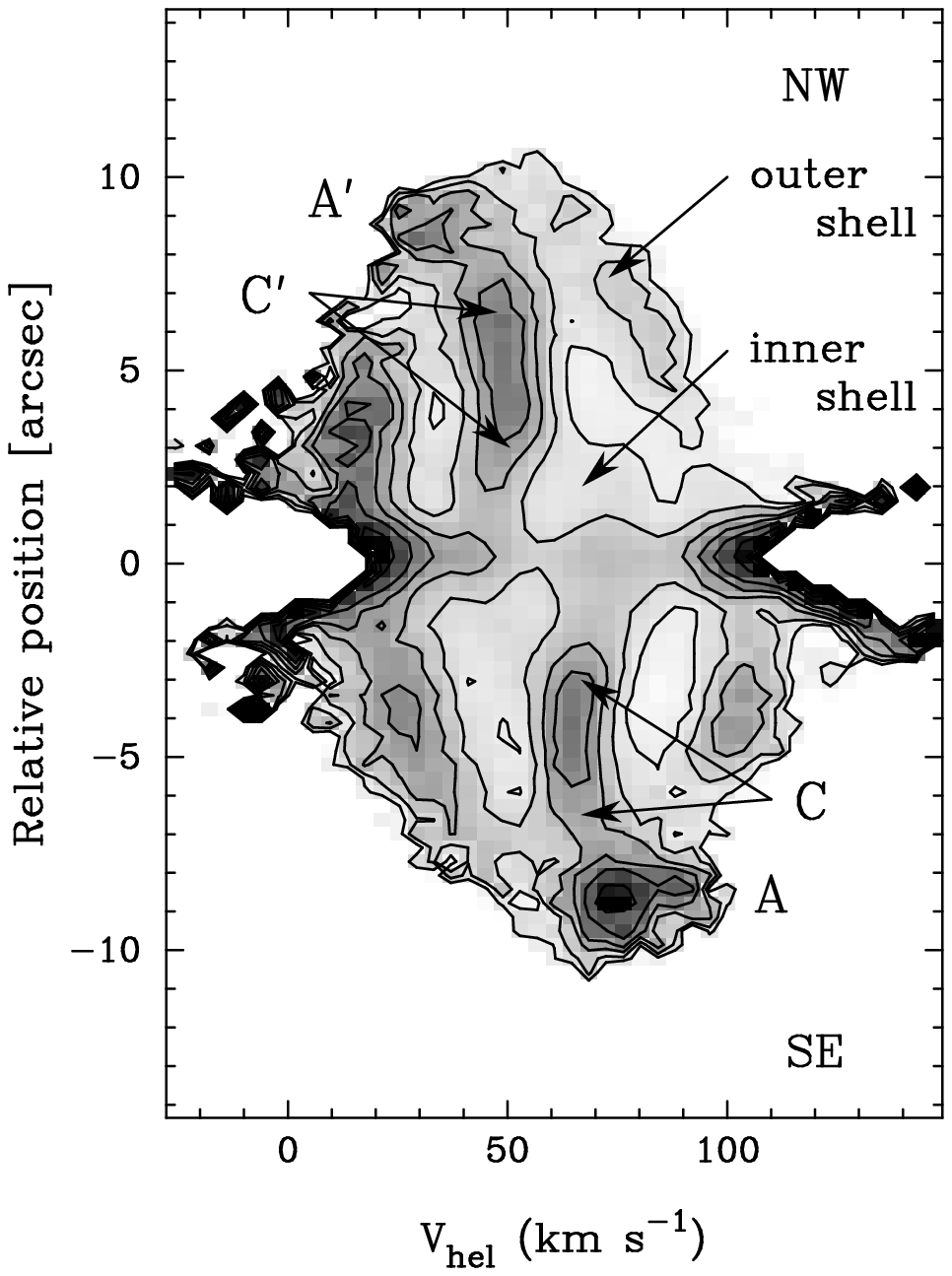}
\end{figure}

\vspace*{10.5cm}
{\bf Figure 5.--}
Negative grey-scale representation of the [N~{\sc ii}] to 
H$\alpha$ emission-line ratio map at PA~{315\arcdeg}. 
The [N~{\sc ii}] enhanced emission regions are shown as dark grey. 
The different morphological features have been labelled. 
Note that a mask was applied to the stellar (ratio $\sim1$) 
continuum.

\end{document}